\documentclass[a4paper,12pt]{article}
\usepackage{graphicx}
\usepackage{amsfonts}
\usepackage{amssymb}
\usepackage{amsfonts}
\usepackage{amsmath}
\usepackage{pifont}
\usepackage{graphicx}
\usepackage{verbatim}
\usepackage{cite}
\usepackage[utf8x]{inputenc}
\def \be {\begin{eqnarray}}
\def \ee {\end{eqnarray}}
\def \ll {{\cal L}}
\def \lc {\left(}     

\def \rc {\right)}     

\def\pa{\partial}  

\begin{document}

\markboth{A. L. dos Santos and D. Hadjimichef}
{Searching for a dark matter coupling to the Standard Model ...
}

\title{\bf Searching for a dark matter coupling to the Standard Model with a Stueckelberg extension}
\author{A. L. dos Santos$\thanks{%
e-mail: alunkes.s@if.ufrgs.br}$, D. Hadjimichef $\thanks{%
e-mail: dimiter.hadjimichef@ufrgs.br}$ \and Instituto de F\'{\i}sica, Universidade Federal do Rio Grande do Sul \\ 
Porto Alegre, RS, Brazil}

\maketitle

\begin{abstract}
We investigate a double extension to the Standard Model (SM). 
A first extension introduces, via   minimal coupling, a massive  $Z'$ boson. This enlarged SM
is coupled to a dark matter  sector  through the Stueckelberg mechanism by a 
$A'$ boson.
 
However, the $A'$ boson does not interact directly with the SM fermions.
In our study, we found  that the $A'$ is a massless photon-like particle in dark sector.
Constraints on the mass for $Z'$ and corrections to $Z$ mass are obtained.
\end{abstract}


\section{The Lagrangian density of the model}	
\noindent 
The  SM electro-weak gauge group is extended from  $SU(2)_L\otimes U(1)_Y$
to $SU(2)_L\otimes U(1)_Y\otimes U(1)_C\otimes U(1)_X$. 
The model lagrangian density that represents this gauge group is written  
in a ge\-ne\-ral form 
as a sum of 

\be
\ll = \ll_{\rm \,eSM}+\ll_{\rm St},
\label{lag}
\ee
where the enlarged Standard Model (eSM) part is defined by   

\begin{eqnarray}
{\cal L}_{\rm\, eSM}&=& -\frac{1}{4} W^{a}_{\mu\nu}W^{a \mu\nu}-\frac{1}{4}B_{\mu\nu}B^{ \mu\nu}
-\frac{1}{4}C_{\mu\nu}C^{ \mu\nu}
\nonumber\\
&&-D_{\mu} \Phi^\dagger D^\mu \Phi -V(\Phi^\dagger \Phi)+ i\bar{\psi}\gamma^\mu D_{\mu} \psi,
\end{eqnarray}

which represents a  modified SM, obtained by introducing a new $U(1)$ boson field   $C_{\mu}$ 
and the field tensor $C_{\mu\nu}=\pa_\mu\,C_\nu-\pa_\nu\,C_\mu$.
 This field is coupled to
the other SM bosons in the covariant derivative  \cite{zprime} 

\be
\label{derivada}
D_{\mu}=\partial_{\mu} + i g_2 \frac{\tau^a}{2} W^{a}_{\mu}+i g_Y \frac{Y}{2} B_{\mu}+i g_c \frac{Y}{2} C_{\mu}
\,,
\ee

where  $\Phi$ is the usual Higgs field.

The Stueckelberg dark sector lagrangian is written as \cite{stueckelberg}
\be
{\cal L}_{\rm St}&=&-\frac{1}{4}X_{\mu\nu}X^{ \mu\nu} 
-\frac{1}{2}(\partial_\mu \sigma +M_1 C_\mu+M_2 X_\mu)^2
+{\cal L}_{\chi}
,
\label{dm}
\ee
 
where the new $U(1)$ vector boson in the dark sector is $X_\mu$ with its 
 field tensor $X_{\mu\nu}=\pa_\mu\,X_\nu-\pa_\nu\,X_\mu$. The second term
in (\ref{dm}) is the Stueckelberg coupling term between the 
two  boson fields $C_\mu$ and $X_\mu$ via an axial pseudo-scalar $\sigma$ field.  
The $\sigma$ field  is   unphysical and decouples from all fields 
after gauge fixing.
 The last term ${\cal L}_{\chi}$  is a general fermion term of the dark sector.                                          
This type of  model was first proposed by Ref. \cite{nath1}, but with just one dark field, 
and was a\-pplied as well in Ref. \cite{taiwan}. 
Using this full Lagrangian (\ref{lag}) and taking into account just the terms that contribute 
to the bosons masses, we obtain 

\be
{\cal L}^{'}=\dots -\frac{1}{2}
V^{\mu\,T}\,
 M^2\, V_\mu
\,,
\ee

where $V^{\mu\,T}=\left(X^\mu, C^\mu, B^\mu, W^{3 \mu} \right)$ and the mass matrix is

\be
M^2=
\lc 
\begin{array}{cccc}\label{matriz}
 M_2^2 & M_1 M_2 & 0 & 0 \\
 M_1 M_2 & M_1^2+\frac{g_c^2 v^2}{4} & \frac{1}{4} g_c g_Y v^2 & -\frac{1}{4} g_2
   g_c v^2 \\
 0 & \frac{1}{4} g_c g_Y v^2 & \frac{g_Y^2 v^2}{4} & -\frac{1}{4} g_2 g_Y v^2 \\
 0 & -\frac{1}{4} g_2 g_c v^2 & -\frac{1}{4} g_2 g_Y v^2 & \frac{g_2^2 v^2}{4}
\label{mass}
\end{array}
\rc.
\ee

It is interesting to note that a new orthogonal matrix   $O$
can be obtained such that it diagonalizes  (\ref{mass})   
and transforms the basis from the unphysical  fields $X_\mu, C_\mu, B_\mu, W^3_\mu$ 
into  the real fields $A^{\prime}_\mu, Z'_\mu,Z_\mu, A_\mu$.

\section{Mass Determination}
After diagonalizing the matrix (\ref{matriz}) we obtain $4$  mass eigenvalues. They are

{\small
\be
 M_{Z'}^2,M_Z^2&=&\frac{1}{8} \left[ v^2\left(g_2^2+g_c^2+g_y^2\right)+4 M_1^2+4
   M_2^2 \pm\,\Delta \right]
\nonumber\\
M_\gamma^2 &=& 0
\,\,\,\,\,\,\,\,;
\,\,\,\,\,\,\,\,
M_{A'}^2 = 0.
\ee
where
\be
\Delta=\sqrt{8 M_1^2 \left(4 M_2^2-
v^2\left[g_2^2-g_c^2+g_y^2\right]\right)+\left(v^2 \left[g_2^2+g_c^2+g_y^2\right]-4
   M_2^2\right)^2+16 M_1^4}\,.
\nonumber
\ee
}

We identify the masses with the $Z$ and $Z'$ bosons masses.
One of the zero eigenvalues can be associated, as usual, to the photon and the other one to a new massless boson, 
which we shall call a {\it dark photon} $A^{\prime}$.

\begin{figure}[h]
\centering
\includegraphics[width=10cm]{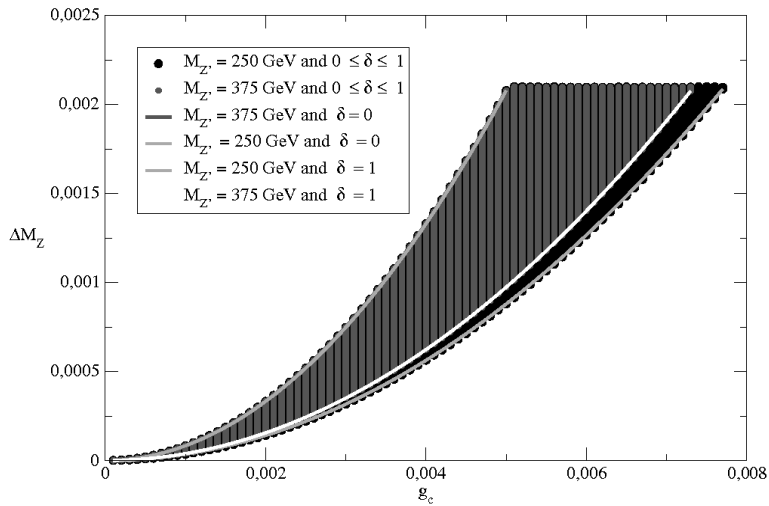}
\caption{A set of values permited to the coupling $g_c$. }
\label{f1}
\end{figure}

The $Z$ mass can be expanded  in terms of $g_c$, this introduces a correction that rises this mass value 
by small quantity $\epsilon$ defined as
{\small
\be
\epsilon = \frac{M_C^2 \left[\left(\delta^2+1\right) \left(M_W^2+M_Y^2\right)-M_{Z'}^2\right]}{\left(\delta^2+1\right)
\left(M_W^2+M_Y^2-M_{Z'}^2\right)}-\frac{\delta^2 M_C^4 M_{Z'}^2}{\left(\delta^2+1\right)^2
\left(M_W^2+M_Y^2-M_{Z'}^2\right)^2} 
\ee
} 

where
\be
\delta=\frac{M_1}{M_2}
\,\,\,\,\,\,;\,\,\,\,\,\,
M_C = \frac{v g_c}{2}
\,\,\,\,\,\,;\,\,\,\,\,\,
M_Y = \frac{v g_Y}{2}\,,
\ee
 $M_W = M_Z\sqrt{1-\sin^2{\theta_W}}$ and $\sin^2{\theta_W} = 0.231$ \cite{pdg}. 
We can compare this correction with the uncertainty in the value of $Z$ mass, $M_Z=91.1876 \pm 0.0021$GeV \cite{pdg}. 
For parameter ajustments,  we set the $Z'$ mass 
in two values \cite{nath1}: $M_{Z'}=250$ GeV and $M_{Z'}=375$GeV, then vary the $\delta$ parameter over the range
$0 \leq \delta \leq 1$. The possible values for 
these parameters are shown in the figure \ref{f1}.


The possible values for $g_c$ are acceptable considering that this range of values is smaller than the usual values for $g_2$ and $g_y$, turning the term associated with the field $C_\mu$ very small, as  expected.
The next steps are to calculate decays and annihilations of particles from dark side into known particles.

\section*{Acknowledgments}
This work was supported by CNPq.


\begin{thebibliography}{0} 

\bibitem{zprime} P. Langacker, {\it Rev. of Mod. Phys.} {\bf 81}, 3 (2009).

\bibitem{stueckelberg} E. C. G. Stueckelber, {\it Helv. Phys. Acta} {\bf 11}, 225 (1938); H. Ruegg and M. Ruiz-Altaba, {\it Int. Jour. of Mod. Phys. A} {\bf 19}, 3265 (2004).

\bibitem{nath1} B. K\"ors and P. Nath,{\it Phys. Lett. B} {\bf 586}, 366 (2004); and {\it JHEP} {\bf 07}, 069 (2005).

\bibitem{taiwan} K. Cheung Tzu-C. Yuan, {\it JHEP} {\bf 03}, 120 (2007); Y. Zhang, S. Wang and Q. Wang, {\it JHEP} {\bf 03}, 047 (2008); D. Feldman, Z. Liu, P. Nath and G. Peim, {\it Phis. Rev. D} {\bf 81}, 095017 (2010).

\bibitem{pdg} K. Nakamura {\it et al}, {\it Jour. of Phys. G} {\bf 37}, 075021 (2010).

\end{thebibliography}
\end{document}